\pdfoutput=1
\documentclass[11pt,a4paper,final,bibliography=leveldown]{iopart}
\usepackage{graphicx}
\usepackage[breaklinks=true,colorlinks=true,linkcolor=blue,urlcolor=blue,citecolor=blue]{hyperref}

\graphicspath{{figures/}}

\newcommand{\LiBi}{$^{209}\rm{Bi}^{80+}$}
\newcommand{\HBi}{$^{209}\rm{Bi}^{82+}$}
\newcommand{\LIBELLE}{{\bf LI}thium-like {\bf B}ismuth {\bf E}xcitation by {\bf L}aser {\bf L}ight at the {\bf E}SR}

\begin{document}
\title{Lifetimes and $g$-factors of the HFS states in H-like and Li-like bismuth}

\author{Volker~Hannen$^1$, Jonas~Vollbrecht$^1$, Zoran~Andelkovic$^2$, Carsten~Brandau$^{2,3}$, 
Andreas~Dax$^4$, Wolfgang~Geithner$^2$, Christopher~Geppert$^{5,6}$, Christian~Gorges$^{5,6}$, 
Michael~Hammen$^{6,7}$, Simon~Kaufmann$^5$, Kristian~K\"onig$^5$, Yuri~A.~Litvinov$^2$, Matthias~Lochmann$^5$,
Bernhard~Maa\ss{}$^5$, Johann~Meisner$^8$, Tobias~Murb\"ock$^9$, Rodolfo~S\'{a}nchez$^2$, Matthias~Schmidt$^8$,
Stefan~Schmidt$^5$, Markus~Steck$^2$, Thomas~St\"ohlker$^{2,10,11}$, Richard~C.~Thompson$^{12}$, 
Christian~Trageser$^3$, Johannes~Ullmann$^{1,5}$, Christian~Weinheimer$^1$ and Wilfried N\"ortersh\"auser$^5$}

\address{$^1$Institut f\"ur Kernphysik, Westf\"alische Wilhelms-Universit\"at M\"unster, Germany}
\address{$^2$GSI Helmholtzzentrum f\"ur Schwerionenforschung, Darmstadt, Germany}
\address{$^3$I.Physikalisches Institut, Justus-Liebig-Universit\"at Gie\ss{}en, Germany}
\address{$^4$Paul Scherrer Institut, Villigen, Switzerland}
\address{$^5$Institut f\"ur Kernphysik, Technische Universit\"at Darmstadt, Germany}
\address{$^6$Institut f\"ur Kernchemie, Johannes Gutenberg-Universit\"at Mainz, Germany}
\address{$^7$Helmholtz Institut Mainz, Johannes Gutenberg-Universit\"at Mainz, Germany}
\address{$^8$Physikalisch-Technische Bundesanstalt, Braunschweig, Germany}
\address{$^9$Institut f\"ur Angewandte Physik, Technische Universit\"at Darmstadt, Germany}
\address{$^{10}$Helmholtz Institut Jena, Germany}
\address{$^{11}$Institut f\"ur Optik und Quantenelektronik, Jena, Germany.}
\address{$^{12}$QOLS Group, Department of Physics, Imperial College London, London, UK.}

\ead{hannen@uni-muenster.de}

\begin{abstract}
The LIBELLE experiment performed at the experimental storage ring (ESR) at the GSI Helmholtz Center for Heavy Ion Research in Darmstadt, Germany, has successfully determined the ground state hyperfine (HFS) splittings in hydrogen-like (\HBi) and lithium-like (\LiBi) bismuth. The study of HFS transitions in highly charged ions enables precision tests of QED in extreme electric and magnetic fields otherwise not attainable in laboratory experiments. Besides the transition wavelengths the time resolved detection of fluorescence photons following the excitation of the ions by a pulsed laser system also allows the extraction of lifetimes of the upper HFS levels and g-factors of the bound $1s$ and $2s$ electrons for both charge states. 
While the lifetime of the upper HFS state in \HBi\ has already been measured in earlier experiments, an experimental value for lifetime of this state in \LiBi is reported for the first time in this work. 
\end{abstract}



\section{Introduction}
Highly charged ions provide a testing ground for QED calculations in extreme electric (up to $10^{16}$\,V/cm) and magnetic (up to $10^4$\,T) fields~\cite{Sha18, Per97} that cannot be created in the laboratory with conventional methods (like lasers and superconducting magnets). Especially interesting in this context is the study of hyperfine transitions in hydrogen-like and lithium-like ions. By taking the so-called specific difference~\cite{Sha01} of the HFS splittings of both charge states, it is possible to cancel nuclear-structure effects (most notably the Bohr-Weisskopf effect caused by the nuclear magnetization distribution) whose uncertainty would otherwise obscure the QED corrections under study.
The LIBELLE\footnote{\LIBELLE} experiment was for the first time able to measure H-like and Li-like HFS splittings in the same isotope~\cite{Ull17} with sufficient precision to allow for a test of bound state QED calculations. 
While the prime target of the experiment was the determination of the transition energies, the time-resolved acquisition of the fluorescence photons emitted during the de-excitation of the ions also allowed us to extract lifetimes of the upper HFS states and, together with the measured transition energies, to calculate $g$-factors of the bound $1s$ electron for H-like bismuth and of the $2s$ electron for Li-like bismuth.
The following sections will provide details on the experimental procedure and analysis of the data required to extract the aforementioned quantities. 
\section{LIBELLE Experiment}
\begin{figure}[h]
 \centering
 \includegraphics[width=\textwidth]{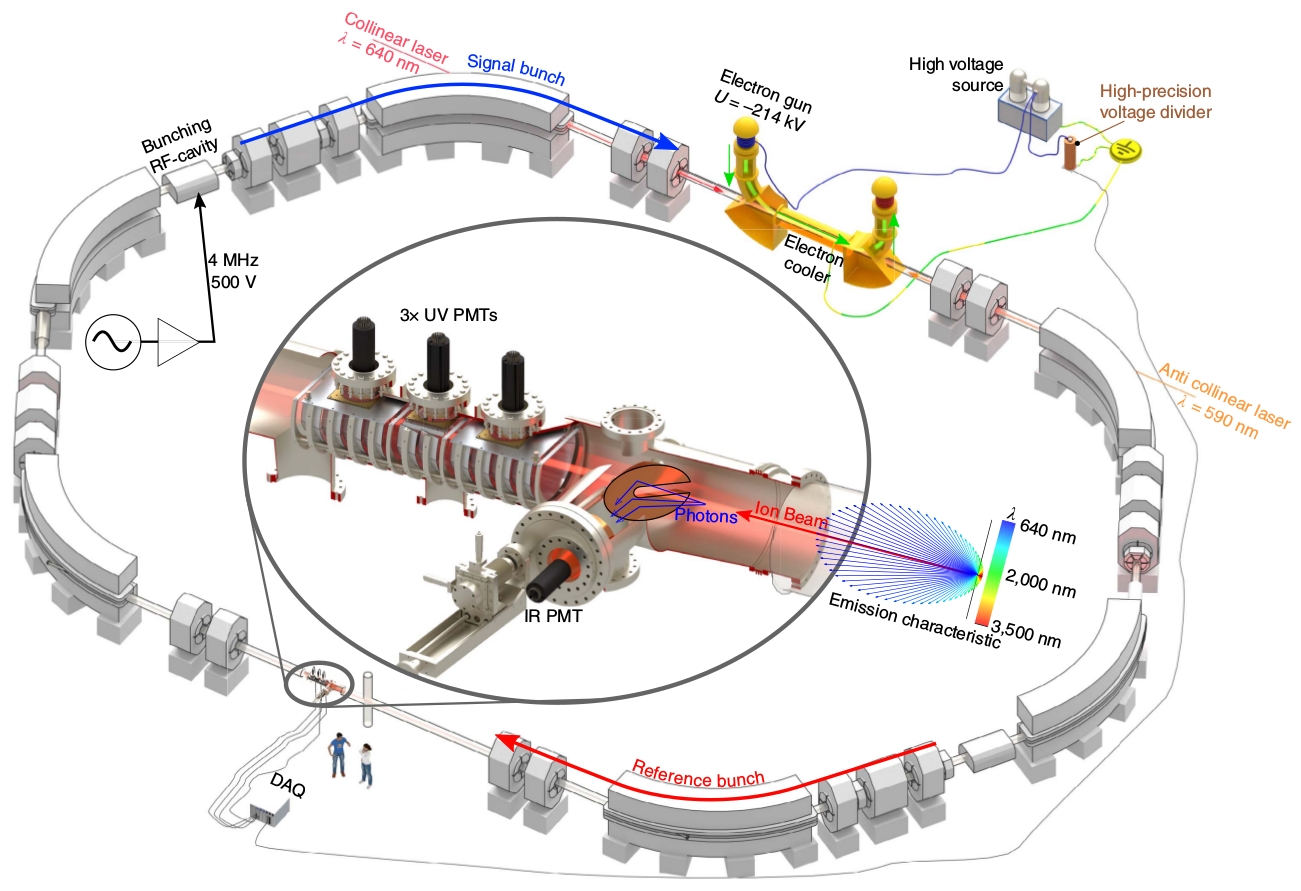}
 \caption{Schematic view of the LIBELLE experiment at ESR (reprinted from~\cite{Ull17} with permission).}
 \label{fig:libelle}
\end{figure}

The experimental setup of the LIBELLE Experiment has been explained in detail in several publications~\cite{Ull17, San17, San15, Ull15, Vol15, Loc14, Nor13} with a focus on the extraction of the wavelengths of the H-like and Li-like HFS transitions. The experiment proceeded in two incarnations with an initial beam-time in 2011 that succeeded for the first time to observe the ground state HFS transition in Li-like bismuth in a laser spectroscopy experiment and a second beam-time in 2014 that featured a greatly increased accuracy in the determination of the level splittings by continuously monitoring the electron cooler voltage with a high-precision HV divider provided by the National Metrology Institute of Germany (PTB) and an improved DAQ system that also allowed us to obtain high statistics data on the lifetime of the two HFS states.\\
Figure~\ref{fig:libelle} gives an overview of the experimental setup.
In the experiment two bunches of either H-like (\HBi) or Li-like (\LiBi) ions are revolving in the Experimental Storage Ring (ESR), located at the GSI facility, at a velocity of $\beta \approx 0.71$. The ion bunches are created 
after the injection of ions into the ring using a radio frequency (RF)-cavity operated at the second harmonic of the revolution frequency ($\approx 2$\,MHz) with a nominal amplitude of 500\,V. 
The stored ions are cooled by means of an electron cooler operated at a voltage near -214\,kV. A tunable pulsed laser system with a repetition rate of 30\,Hz is synchronized with the revolution frequency of the ions and used to illuminate one of the two bunches (called 'signal' bunch in the following). The other ion bunch (called 'reference' bunch) is not illuminated and is used to determine the experimental background. \\
For H-like bismuth the laser beam is injected in a counter propagating manner, shifting the wavelength of the laser from 590\,nm in the laboratory frame to 244\,nm wavelengths in the rest-frame of the stored ions. 
Correctly tuned, the laser will excite the ions to the upper HFS state from which they will subsequently decay with a mean lifetime of about 567\,$\mu$s~\cite{Win98,Win99} in the laboratory frame. 
As the lifetime is long compared to the revolution period of the ions in the ring, fluorescence photons from the de-excitation are created along the whole circumference of the ESR and can therefore also be detected on the opposite side of the storage ring, using a suitable mirror system and photomultipliers (PMTs) situated on top of UV transparent windows in the beam-line (leftmost 3 PMTs visible in the inset of figure~\ref{fig:libelle})~\cite{See98,San17}.\\
For Li-like bismuth the laser beam is injected in a co-propagating manner, shifting the wavelength of the laser from 640\,nm in the laboratory frame to 1554\,nm in the rest-frame of the stored ions. 
The theoretically predicted lifetime of the $F=5$ state in \LiBi of about 118\,ms~\cite{SST98} at $\beta=0.71$ is even longer and allows us again to detect fluorescence photons opposite to the electron cooler after resonant excitation with the laser.
For that purpose a specially developed movable parabolic mirror system~\cite{Han13} (see right part of inset in figure~\ref{fig:libelle}) has been used. The need to develop a special detection system for the Li-like transition arises from the low signal rate connected to the long lifetime of the state and the long wavelength of the fluorescence photons that are outside the sensitive range of most PMTs.
It is thus necessary to collect the most forward emitted photons that undergo the largest Doppler shift to shorter wavelengths and at the same time make use of the Lorentz boost of the emission characteristics to smaller polar angles.
\section{Lifetime measurement}
Besides the transition wavelengths the time resolved detection of the fluorescence photons following the excitation of the ions by the pulsed laser system allows us to extract the lifetimes of the upper HFS states in both \HBi\ and \LiBi.
\subsection{Method}
\begin{figure}[h]
  \centering
  \includegraphics[width=\textwidth]{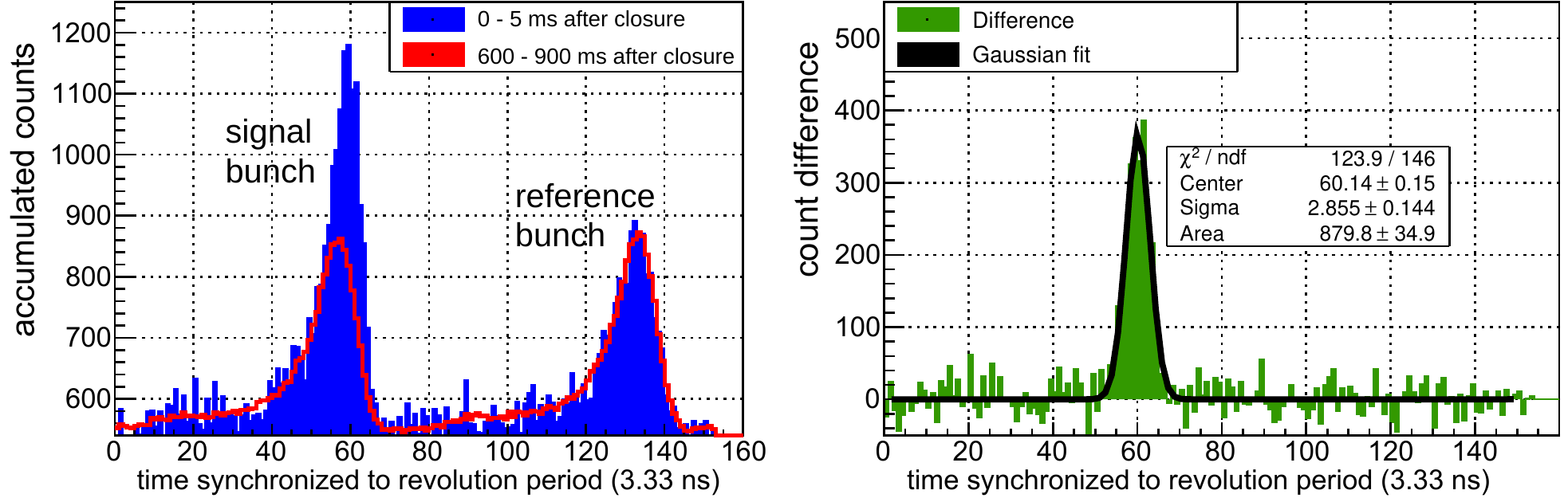}
  \caption{Left, blue histogram: signal and reference peaks for the HFS transition in Li-like bismuth accumulated during the first 5\,ms after shutter closure. The red histogram is obtained by averaging over a time range from 600\,ms to 900\,ms after shutter closure, where the excited ions have almost completely decayed. Right: difference of the two histograms shown on the left, fitted with a Gaussian function.}
  \label{fig:diff}
\end{figure}

\begin{figure}[h]
  \centering
  \includegraphics[width=0.8\textwidth, height=5cm]{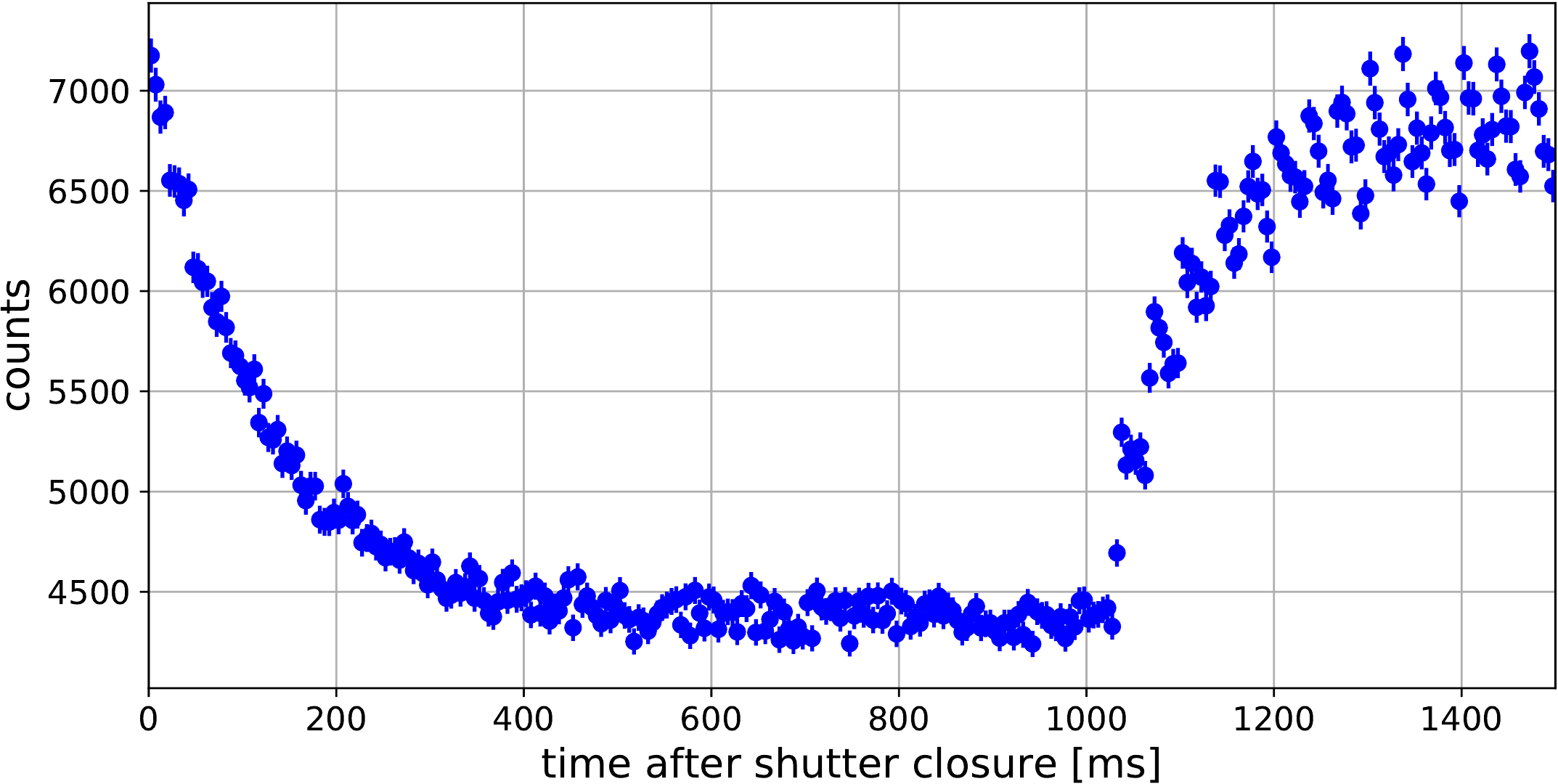}
  \caption{Fluorescence signal observed from Li-like bismuth (\LiBi) after resonant excitation of the M1 HFS transition. The shutter to the laser system is re-opened around 1000\,ms after closure.}
  \label{fig:li_like_data}
\end{figure}

For that purpose, the laser is fixed to the resonance wavelength and the rate of fluorescence photons is measured as a function of time after the laser pulses. While the time interval of 33.3\,ms between individual laser pulses is sufficiently long for the excited HFS state in H-like bismuth to decay completely, the longer lifetime of the upper HFS state in Li-like Bi of about 118\,ms at $\beta=0.71$ requires a shutter system to periodically block the laser for several lifetimes, allowing the observation of a sufficiently large part of the decay curve to obtain a good fit of the lifetime in the laboratory frame $\tau_{\rm lab}$.\\
For the analysis the acquired signals are accumulated in timing bins with a width of 25\,$\mu$s for  H-like ions and 5\,ms for Li-like ions. For each of these bins, histograms of the number of counts as a function of time synchronized to the revolution period of the ions are extracted (see figure~\ref{fig:diff}, left plot). 
In such a display, signals from the two ion bunches appear as peaks with one-sided tails resulting from an afterglow of excited residual gas molecules after passage of the bunches. 
The bin width of these histograms is given by the time resolution of the time-to-digital converter (TDC) which operates with a 300\,MHz clock. 
The left hand side of figure~\ref{fig:diff} shows in blue the accumulated counts for Li-like ions during the first 5\,ms after excitation by the laser on resonance and subsequent closure of the shutter system. The signal peak (left) clearly exhibits a significantly higher number of events caused by fluorescence from the decay of the excited ions together with background processes, while the reference peak (right) contains background events only. 
Several lifetimes after shutter closure, the population of the upper HFS state has almost completely returned to the ground state and signal and reference peaks now both exhibit roughly the same number of background counts (red histogram, obtained by summing up the individual 5\,ms histograms in the time interval between 600\,ms and 900\,ms and dividing by their number).
The right hand side in figure~\ref{fig:diff} displays the difference of the blue and red histograms that corresponds to photons from the laser-excited HFS states only. This plot allows the determination of the center and width of the fluorescence signal contribution to the overall event display.\\
To extract lifetime curves of the HFS transitions from the bunch histograms obtained for each time bin, the most robust method turned out to be a summation of the signal peak in a $\pm n\sigma$ window around the center of the signal count distribution determined as described above~\cite{Vol16} (for the choice of the actual width of the summation window, see section~\ref{sec:systematics}). 
No background subtraction is performed in the process as especially for the parts of the lifetime curves where most HFS states have already decayed, one deals with the subtraction of two similarly large numbers that only leads to an increase in the statistical uncertainty of the extracted signal counts. 
Instead the background counts are summed together with the HFS signals and lead to a constant offset in the extracted lifetime curves that is treated as a fit parameter in the subsequent analysis. 
Figure~\ref{fig:li_like_data} displays an example for such a lifetime curve obtained with Li-like bismuth ions.
The number of counts is plotted against the time after closure of the shutter required due to the long decay time in the case of Li-like bismuth. In the first 1000\,ms one observes the expected exponential decay of the signal rate together with a constant offset from background events (approximately 4370 in this case). Following that, the shutter is opened for 500\,ms such that the upper HFS state can be populated again by the laser. The displayed data is accumulated over many shutter cycles to obtain sufficiently high statistics. The shutter cycle is synchronized to the timing of the laser pulses and the pulse structure of the laser with 33.3\,ms period can therefore be observed in the rising branch of the data when the shutter is open. 
\subsection{Fit model}
\label{sec:fit}
While fitting lifetime data from the measurements with H-like bismuth ions is straightforward as one is dealing with a simple exponential decay plus background, more effort is required in the Li-like case, where one would also like to extract the information contained in the rising branch of the curve, during which the upper HFS state 
is gradually repopulated by a number of consecutive resonant laser pulses.
The fit-model for the latter case has to take into account the processes of excitation, stimulated emission and spontaneous emission. The rate equation for the population of the upper HFS state $N_2(t)$ reads:
\begin{equation}
 \dot{N_2}(t) = N_1(t) \, B_{12} \, u - N_2(t) \, B_{21} \, u - N_2(t) \, A_{21} \; ,\label{rate:eq}
\end{equation}
with the Einstein coefficients 
\begin{equation}
 A_{21} = \frac{1}{\tau}\hspace{10mm}B_{21} = A_{21} \frac{\lambda^3}{8\pi h}\hspace{10mm}g_1 \,  B_{12} = g_2 \,  B_{21} \;  .
\end{equation}
The coefficient for spontaneous emission $A_{21}$ is the inverse of the mean lifetime $\tau$. $g_1$ and $g_2$ are the number of magnetic sub-states of the ground state ($F = 4$) and the excited state ($F = 5$) of the given HFS transition. $N_1(t)$, $N_2(t)$ are the occupation numbers of the two states, which yield the total number of stored ions $N$ via \mbox{$N=N_1(t)+N_2(t)$}. 
As the typical storage time of the ions in the ring was about 20 minutes, the time dependence of the total ion number over the timescale of a single 1.5\,s shutter cycle can be neglected.
$\lambda$ denotes the wavelength of the transition and $u$ the time averaged spectral energy density of the laser. Equation~\ref{rate:eq} can be re-written as follows:
\begin{equation}
 \dot{N_2}(t) = \left\{\frac{g_2}{g_1}N - \left(\frac{g_2}{g_1} + 1\right) N_2(t)\right\} \, \frac{\lambda^3 u}{8\pi h} \,\frac{1}{\tau} - N_2(t) \,\frac{1}{\tau}
 \label{advanced:eq}
\end{equation}
With the abbreviation $\phi := \frac{\textstyle\lambda^3 u}{\textstyle8\pi h}$ eq.~\ref{advanced:eq} simplifies to 
\begin{equation}
  \dot{N_2}(t) = \underbrace{\frac{g_2}{g_1}N \frac{\phi}{\tau}}_{\textstyle:= a} - 
                \underbrace{\left(\frac{\left(\frac{g_2}{g_1} + 1\right) \phi + 1}{\tau} \right)}_{\textstyle:= b} N_2(t) =  a-bN_2(t) \label{tosolve}
\end{equation}
The differential equation~\ref{tosolve} can be solved using the ansatz 
$N_2(t) = \frac{a}{b} + c \, e^{-b \, t}$, 
where $c$ is a constant, introduced by the boundary conditions at $t=0$ and is therefore given by
$c = N_2(0) - \frac{a}{b}$. Together it yields
\begin{equation}
 N_2(t) = \frac{a}{b} + \left( N_2(0) - \frac{a}{b} \right) \exp\left(-b\, t \right) \label{ExcIons:eq}
\end{equation}
Stimulated emission as well as excitation do not take place during the full measurement interval since the laser is blocked for a defined time. Therefore a distinction is needed, with the two cases 
\begin{enumerate}
 \item $t\leq t_c$: The laser is blocked, therefore only spontaneous emission takes place
\vspace*{-0.3mm}
 \item $t>t_c$: The ions are excited by the the laser and all three processes are taken into account
\end{enumerate}
The corresponding fit function is given by
\begin{equation}
 f(t) = \left\lbrace
             \begin{array}{ll}
               s \cdot N_2(0) \exp \left( -\frac{t}{\tau} \right) + d & {\rm for} \;\; t \leq t_c \\[2mm]
               s \cdot \left[ \frac{a}{b} + \left( N_2(t_c) - \frac{a}{b} \right) \exp\left(-b\, (t-t_c) \right)  \right] + d & {\rm for} \;\; t > t_c  \; ,
             \end{array}
           \right.
\label{eq:extfit}
\end{equation}
where $d$ accounts for a constant background and $s$ is a scaling factor representing further experimental parameters like measurement time and observed solid angle. Equation~\ref{eq:extfit} holds for excitation with a continuous laser beam and does not yet take into account the pulse structure of the laser. Therefore a further modification of the model has been implemented in the fitting code, taking into account that during the rising branch of the lifetime curve only each 6th or 7th bin actually contains a laser shot. 
\subsection{Beam-time 2014}
\begin{figure}[h]
 \centering
 \includegraphics[width=0.8\textwidth, height=55mm]{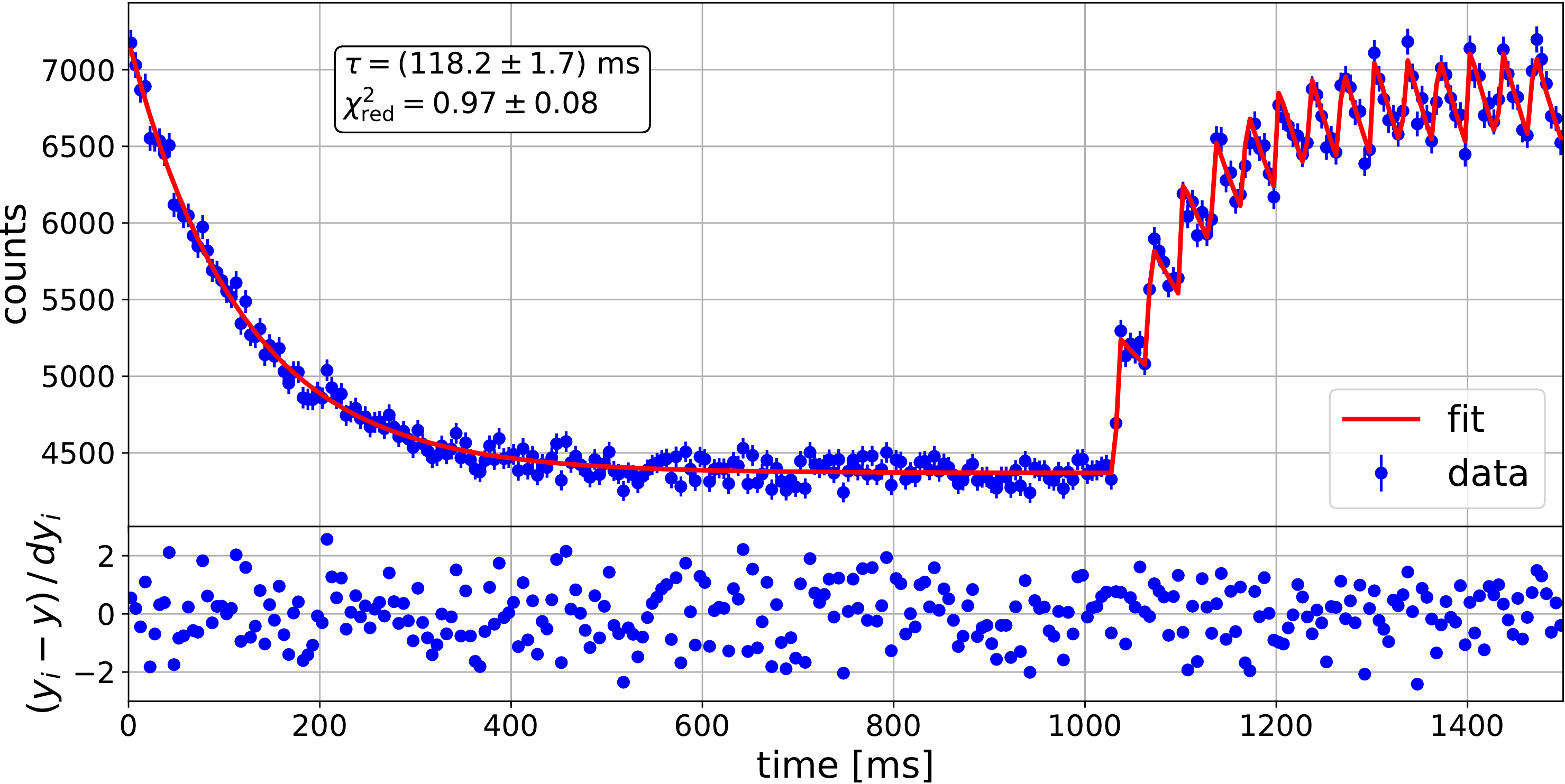}\\
 \includegraphics[width=0.8\textwidth, height=55mm]{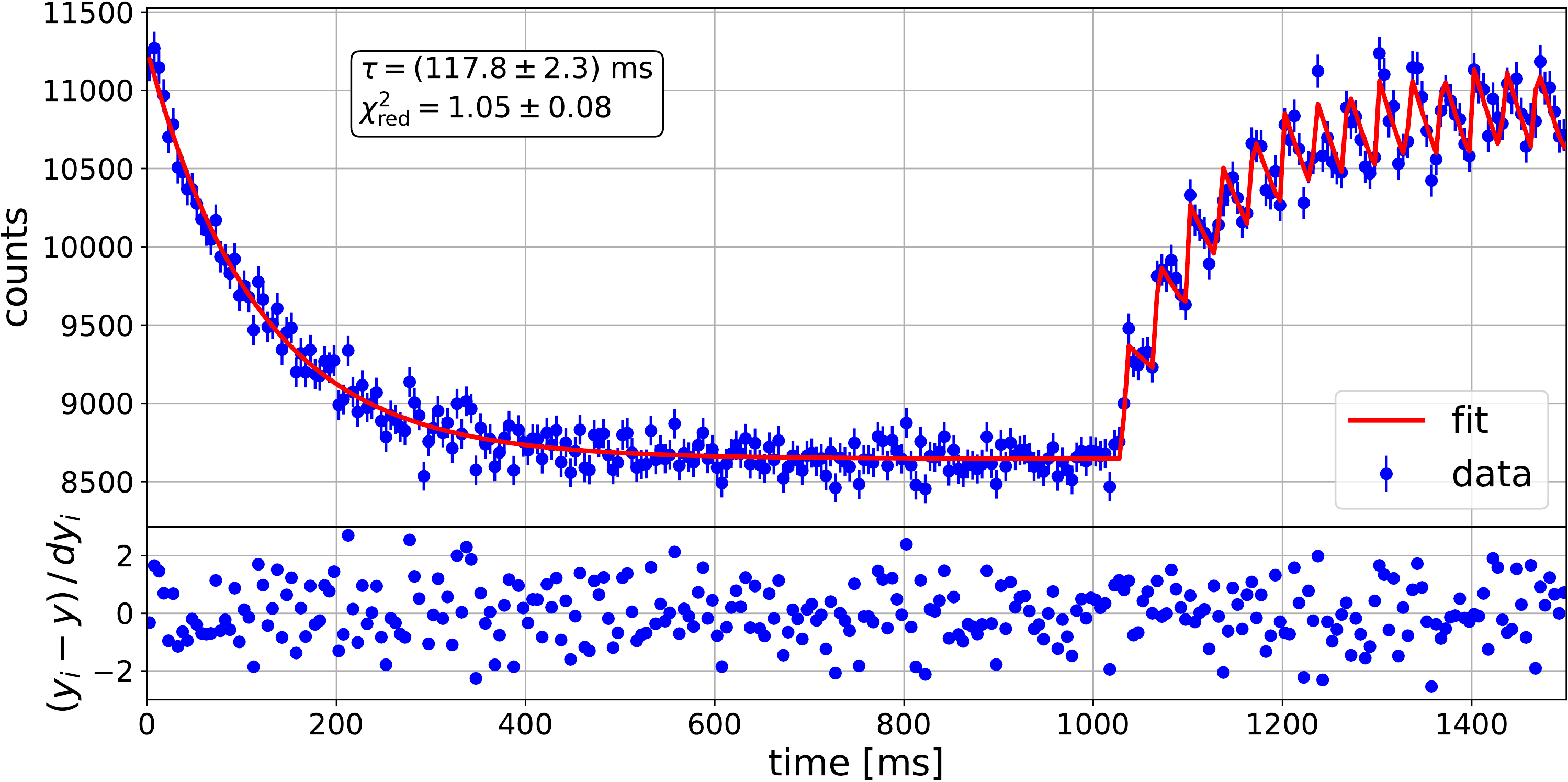}
 \caption{Accumulated data for the lifetime of the upper HFS state in \LiBi taken before (top) and after (bottom) the electron cooler maintenance. The data is fitted with the model described in section~\ref{sec:fit} taking the laser pulse structure into account for times larger 1000\,ms. The lower parts of both plots display the normalized residuals of the fits which are, to good approximation, randomly distributed.}
 \label{fig:li_like_fit}
\end{figure}

\begin{figure}[h]
 \centering
 \includegraphics[width=0.8\textwidth, height=55mm]{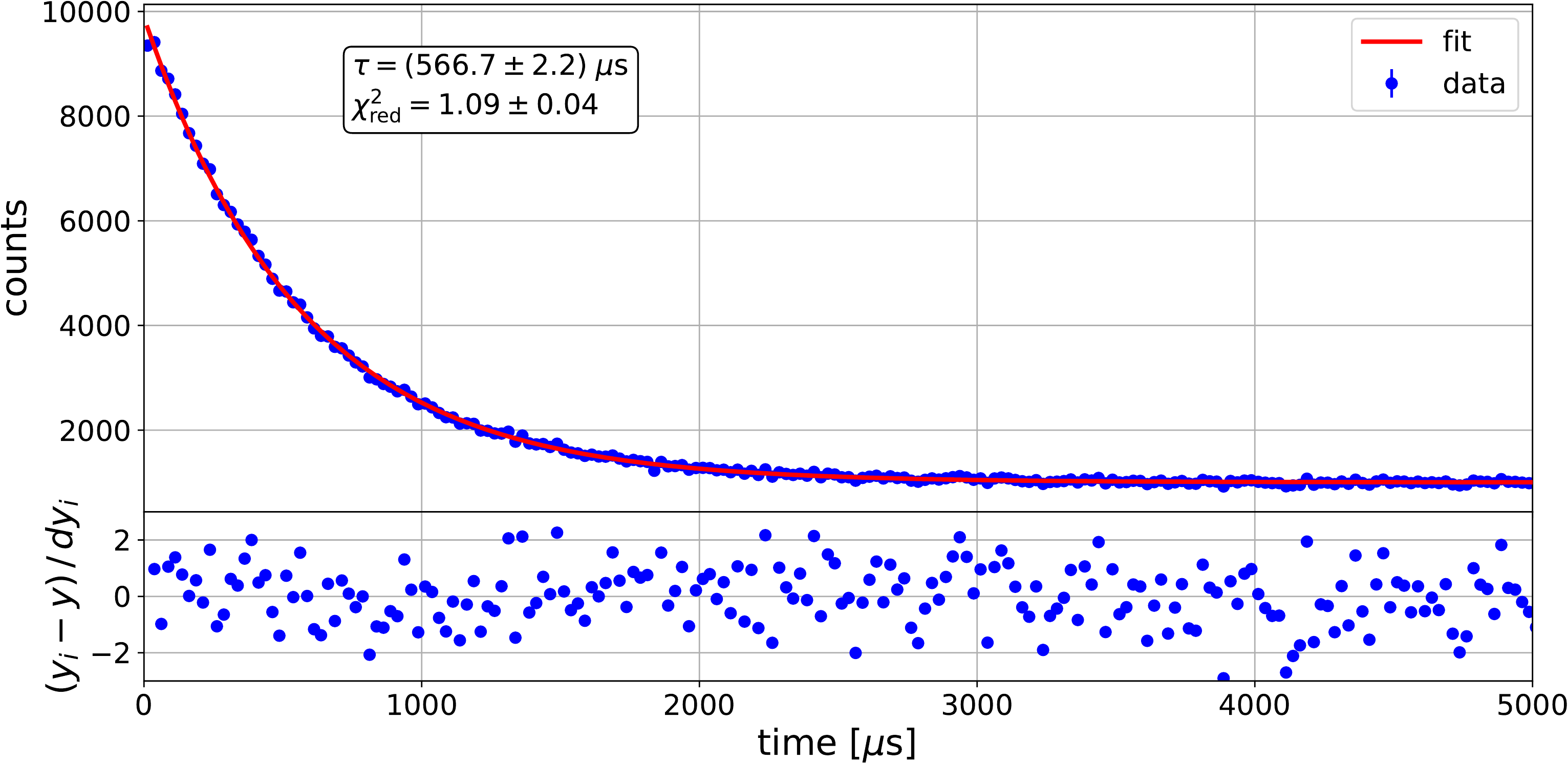}
 \caption{Accumulated data for the lifetime of the upper HFS state in \HBi. The data is fitted by an exponential function with a constant background term.}
 \label{fig:h_like_fit}
\end{figure}

The main part of the LIBELLE beam-time in 2014 was spent on the precise determination of the wavelengths of the HFS transitions in \HBi\ and \LiBi. For that purpose the laser wavelength was scanned repeatedly across the respective transitions and a number of dedicated systematics measurements, among others investigating the influence of the electron cooler current or the buncher amplitude on the observed transition wavelength, were performed~\cite{Ull17, San17}. In addition to this, some runs were recorded in which the laser was fixed to the transition wavelength of either \HBi\ or \LiBi\ (in the latter case with the optical shutter system in operation) to obtain high statistics data on the lifetime of the states. 
\paragraph{Li-like bismuth}
After the first half of the beam-time it was discovered that one of the drift-electrodes inside the electron cooler was not properly connected leading to fluctuations in the revolution frequency of the ion bunches that were then compensated by adapting the electron cooler voltage. After the electrode had been properly grounded, these fluctuations disappeared. At the same time however, the experimental background in the fluorescence detection increased, possibly due to a slight change in the beam properties following the repair action. 
We therefore collect the lifetime measurements for the HFS transition in Li-like bismuth in two datasets, one before and one after the maintenance of the electrode. Figure~\ref{fig:li_like_fit} displays the corresponding
datasets together with results from the fit. The reduced $\chi^2$-values of both fits are consistent with one
indicating that the model provides an adequate description of the data.
The residuals do not show any non-statistical fluctuations and follow a Gaussian distribution around zero.
The extracted lifetimes of both datasets are consistent with each other and with fit results obtained from the 
first 1000\,ms of the data, where only the exponential decay is observed (see table~\ref{tab:li_fit}).
\begin{table}[h]
 \caption{Fit results for Li-like bismuth lifetimes as observed in the laboratory frame before and after the maintenance. Lifetimes were extracted using the model described in section~\ref{sec:fit}, using only the exponentially decaying part of the lifetime curves (0-1000\,ms) or the complete datasets (0-1500\,ms). 
 The uncertainties are purely statistical.} \label{tab:li_fit}
 \centering
 \vspace{1mm}
 \small
 \begin{tabular}{l|c|c}
 Fit interval       & 0 - 1000\,ms & 0 - 1500\,ms \\
 \hline
 before maintenance & $(118.3 \pm 2.1)$\,ms & $(118.2 \pm 1.7)$\,ms \\
 after  maintenance & $(118.5 \pm 3.2)$\,ms & $(117.8 \pm 2.3)$\,ms \\
 \hline
 weighted mean      & $(118.4 \pm 1.8)$\,ms & $(118.1 \pm 1.4)$\,ms \\
 \end{tabular}
\end{table}
\paragraph{H-like bismuth}
For H-like bismuth the relevant measurements were performed after the repair and we are therefore dealing with one dataset only. Figure~\ref{fig:h_like_fit} displays the accumulated data together with the corresponding 
fit result delivering a value for the lifetime of the upper HFS state in \HBi of $(566.7 \pm 2.2)\;\mu$s in the laboratory frame. 
\subsection{Systematic uncertainties}
\label{sec:systematics}
\begin{figure}[h]
 \centering
 \includegraphics[width=0.49\textwidth]{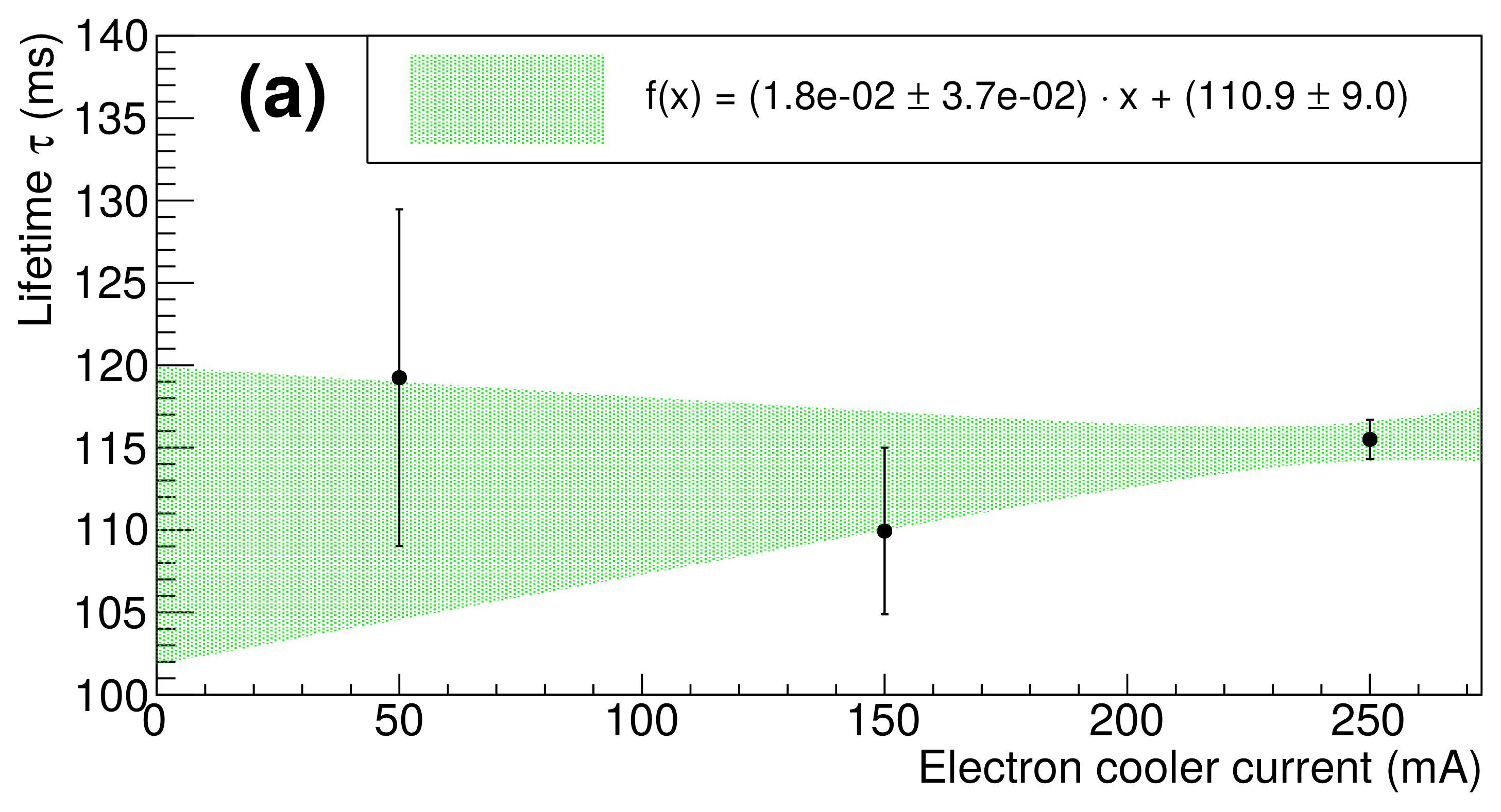}
 \includegraphics[width=0.49\textwidth]{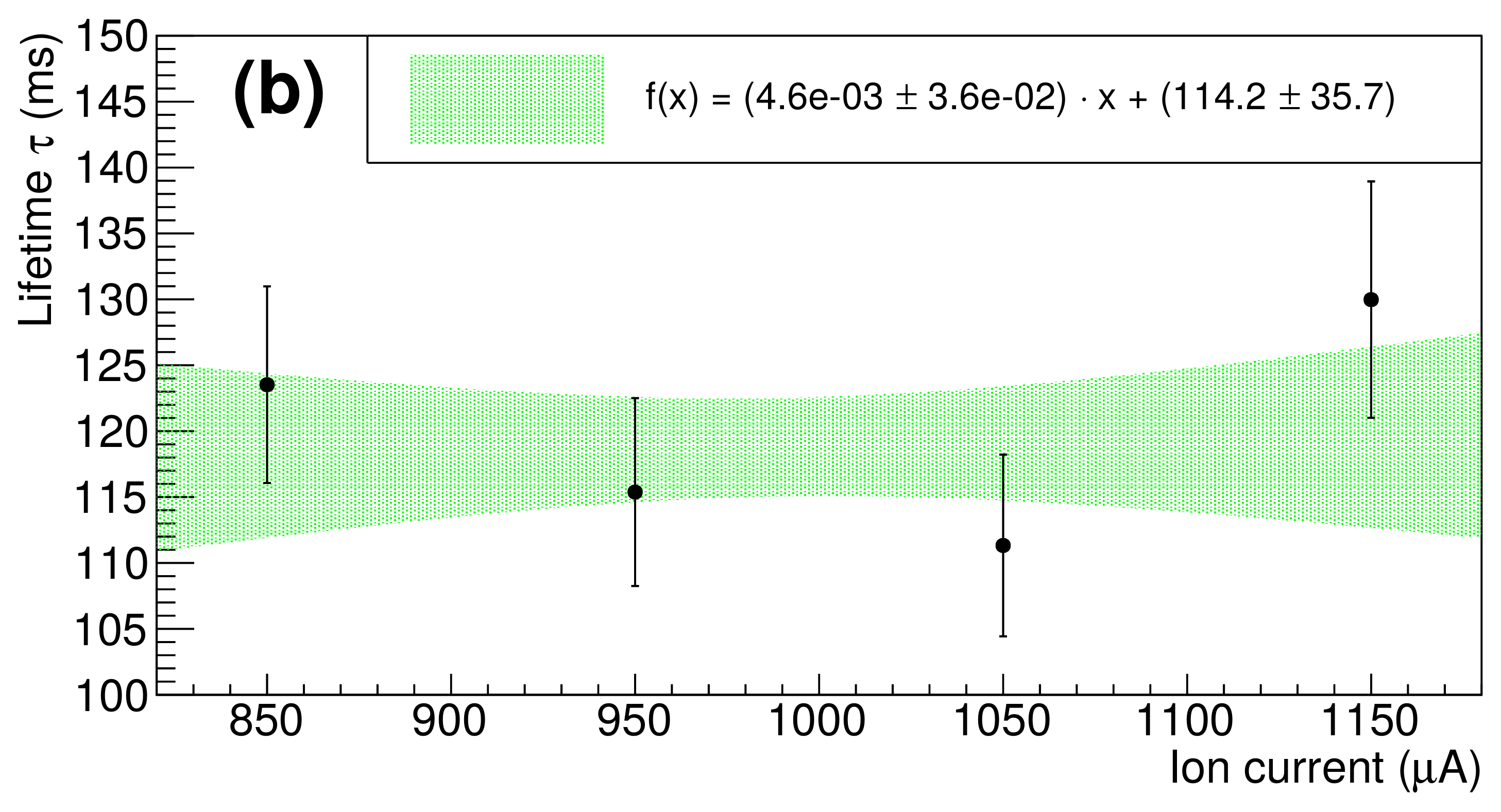}
 \includegraphics[width=0.49\textwidth]{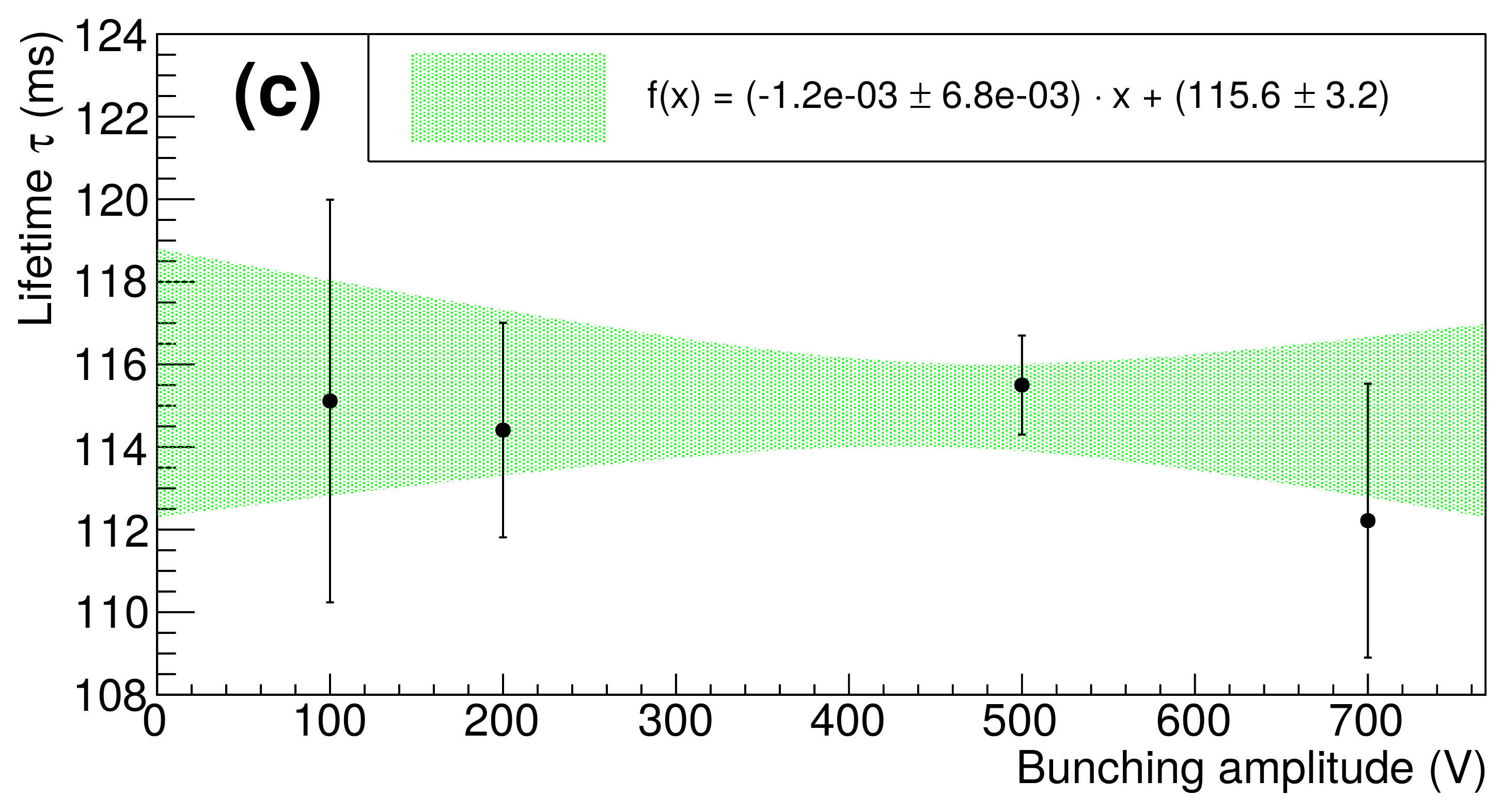}
 \includegraphics[width=0.49\textwidth, height=4cm]{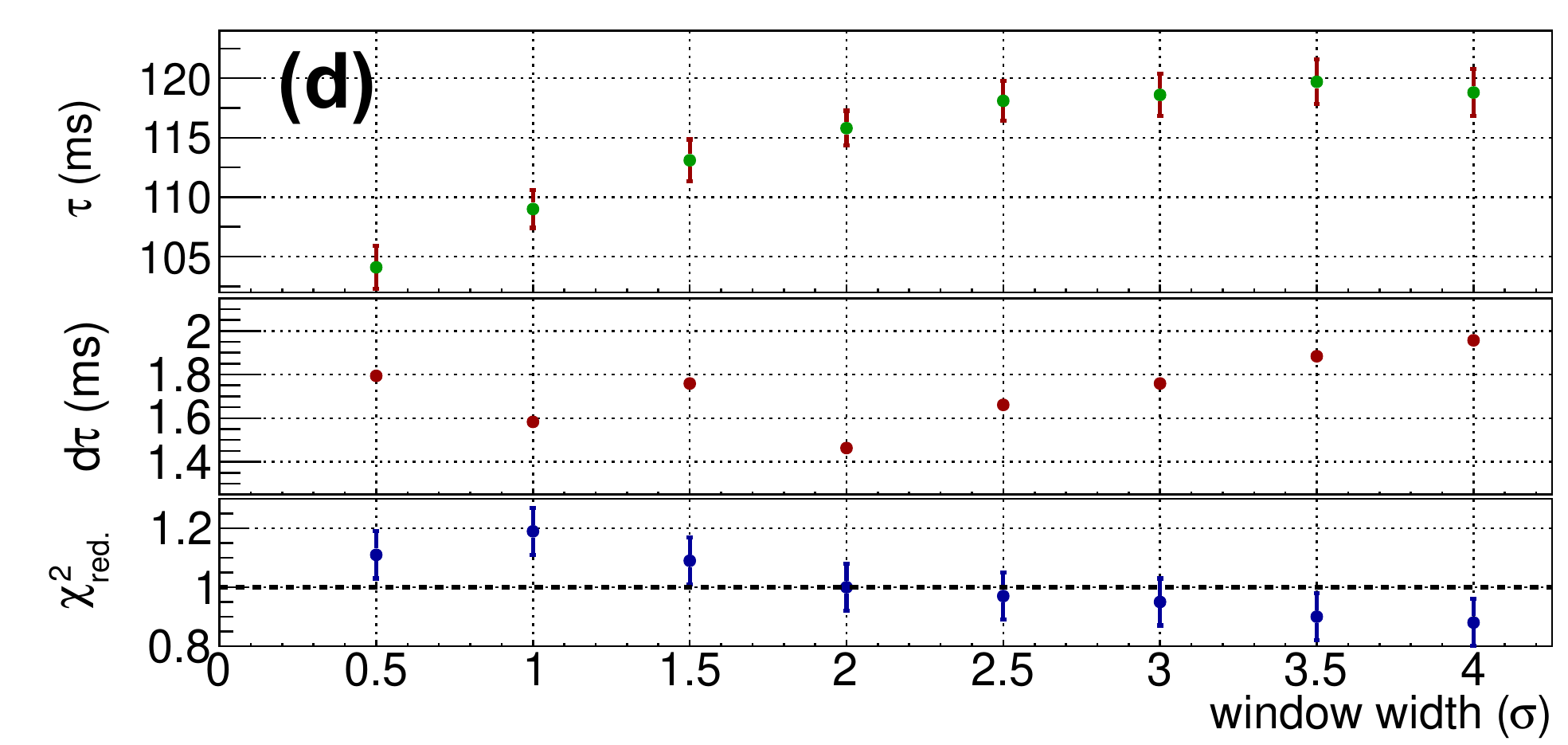}
 \caption{Investigation of systematic uncertainties for the lifetime of the upper HFS state in Li-like bismuth. (a) dependence of the extracted lifetime on electron cooler current, (b) the ion current and (c) the bunching amplitude. Plots (a-c) were produced using a $2\sigma$ summation window. (d) dependence of the extracted lifetime on the width of the summation window used to construct the lifetime curves.}
 \label{fig:li_like_sys}
\end{figure}

A number of measurements were performed to investigate possible systematic effects on the measured wavelengths of the HFS transitions in \HBi and \LiBi~\cite{Ull17, Ull15}. By selecting from these runs only those periods when the laser was on resonance, we can also perform checks of systematic effects on the lifetime of the HFS states, albeit with very restricted statistics. 
Figure~\ref{fig:li_like_sys} displays the results of these investigations for the lifetime of the upper HFS state in Li-like bismuth. 
To investigate possible effects of space charge of the electron beam generated by the cooler or of the ion beam itself, lifetimes have been extracted for different values of the electron cooler current and the ion current (figure~\ref{fig:li_like_sys}, (a) and (b)). Additionally a series of measurements were taken with different amplitudes of the RF buncher voltage that is used to concentrate the ions stored inside the ESR into two bunches (figure~\ref{fig:li_like_sys}, (c)).
The green bands illustrate $1\sigma$ uncertainty intervals of linear fits applied to the corresponding displayed data points.
Within the uncertainties the slopes of all three fits are compatible with zero and, therefore, none of the described measurements displayed a systematic change of the lifetime with the respective experimental parameters. The same findings apply to the systematic tests of the lifetime of the upper HFS state in \HBi. \\
A systematic uncertainty arising in the analysis of the experimental data is given by the chosen width of the analysis window used to sum up the counts in the signal peaks of the individual bunch histograms (see  figure~\ref{fig:diff}). Choosing a too small window leads to shifts in the observed lifetimes as the widths of the signal peaks has been found to slightly fluctuate over the decay period~\cite{Vol16}. Choosing a too large window on the other hand results in an increased fit uncertainty and a lesser quality of the fit demonstrated by a reduced $\chi^2$-value that starts to significantly deviate from one. The effect of a systematic variation of the width of the analysis window from $\pm 0.5\sigma$ to $\pm 4\sigma$ is displayed in figure~\ref{fig:li_like_sys}~(d), for the Li-like dataset before the cooler maintenance. It can be seen that for small window widths the lifetime of the state is underestimated but stabilizes when moving to larger widths. 
The optimum value chosen for the analysis is $\pm 2.5\sigma$, where the lifetime $\tau$ has already reached the plateau, the uncertainty $d\tau$ of the fit value is comparably small and the reduced $\chi^2$ is compatible with one. To estimate the systematic uncertainty connected to this analysis procedure, we calculate the differences between the lifetimes derived with $\pm 2\sigma$, $\pm 2.5\sigma$ and $\pm 3\sigma$ width of the analysis window. The weighted mean of these differences is then taken as systematic uncertainty of the analysis of the given dataset.
The systematic uncertainty for the lifetime of the upper HFS state in \HBi was calculated in the same way.\\
Overall we arrive at the following lifetimes for the upper HFS states in \HBi\ and \LiBi in the laboratory frame:
\begin{eqnarray}
 \tau_{\rm lab}(^{209}\rm{Bi}^{82+})  & = & (566.7 \pm 2.2_{\rm stat.} \pm 1.3_{\rm sys.})\; \mu{\rm s} \\
 \tau_{\rm lab}(^{209}\rm{Bi}^{80+}) & = & (118.1 \pm 1.4_{\rm stat.} \pm 1.3_{\rm sys.})\; {\rm ms} \; .
\end{eqnarray}
%
%
%
%
\subsection{Lifetimes in the rest frame of the ions}
To transform the measured lifetimes from the laboratory frame to the rest frame of the ions we need to know the velocities of the ions. 
In the experiment the ion velocity is determined by the velocity of the electrons in the cooler, which itself can be calculated from the acceleration voltage applied to the cooler and space-charge corrections taking into account a reduction of the potential felt by the electrons due to shielding effects in the electron beam.
%
%
To determine the latter, several measurements were conducted with H-like ion beams during which the electron cooler current was lowered stepwise and the change in velocity of the ion beam was then corrected for by adapting the electron acceleration voltage until the original revolution frequency of the ions (and thus their velocity in the ring) was restored.
From a linear fit to the obtained data points, the space-charge correction was found to be 
$(-0.162 \, \pm \, 0.009)$\,V/mA~\cite{Ull15}.
The average voltage during the lifetime measurements extracted from the datasets and corrected for the space-charge effect is for the case of H-like bismuth found to be $U_{\rm eff}(^{209}\rm{Bi}^{82+}) = (-213878 \pm 9)$\,V and for Li-like bismuth $U_{\rm eff}(^{209}\rm{Bi}^{80+}) = (-213871 \pm 10) $\,V (the latter value has been obtained from the datasets after the cooler maintenance).
Besides the uncertainties of the space charge correction the quoted error bars include the standard deviation of the high voltage measurements from the calculated mean (1.3\,V for \HBi\ and 2.1\,V for \LiBi), the uncertainty of the scale factor of the PTB HVDC voltage divider (13\,ppm~\cite{Hal14}) and uncertainties due to the difference in contact potentials of cathode and drift tubes of the electron cooler ($\leq 3$\,V). For a conservative estimate of the overall uncertainty on the effective cooler voltage $U_{\rm eff}$ these contributions were added linearly.\\
We can then calculate the Lorentz factor using
\begin{equation}
 \gamma = 1 + \frac{−e \, U_{\rm eff}}{m_e \, c^2}
\end{equation}
with the electron charge $e$ and electron rest mass $m_e$ to obtain 
$\gamma (^{209}\rm{Bi}^{82+}) = 1.418548(18)$ and $\gamma (^{209}\rm{Bi}^{80+}) = 1.418534(20)$. 
With these results the measured lifetimes can be transfered into the laboratory frame using $\tau = \tau_{\rm lab}/\gamma$ yielding the results listed in table~\ref{tab:lifetimes}.
\begin{table}[b]
 \caption{Results for the lifetimes of the upper HFS state in \HBi\ and \LiBi\ compared to existing experimental and theoretical data.} \label{tab:lifetimes}
 \vspace{1mm}
 \centering
 \small
 \begin{tabular}{l|c|c}
  & H-like bismuth \HBi & Li-like bismuth \LiBi \\
  \hline
  \hline
  this work & $(399.5 \pm 1.6_{\rm stat.} \pm 0.9_{\rm sys.})\; \mu{\rm s}$ & $(83.3 \pm 1.0_{\rm stat.} \pm 0.9_{\rm sys.})\; {\rm ms}$ \\
  \hline
  H. Winter et al. (exp.)~\cite{Win98,Win99}     & $(397.5 \pm 1.5_{\rm stat.})\; \mu{\rm s}$ & \\
  I. Klaft et al. (exp.)~\cite{Kla94}      & $(351 \pm 16_{\rm stat.})\; \mu{\rm s}$ & \\
  V.M. Shabaev (theo.)~\cite{Sha98}        & $(399.01 \pm 0.19)\; \mu{\rm s}$ & \\
  \hline
  V.M. Shabaev et al. (theo.)~\cite{SST98} & & $(82.0 \pm 1.4)$\,ms \\
  Calculated from theoretical & & $(82.85 \pm 0.61)$\,ms \\
  g-factor~\cite{MSQ08} and wavelength~\cite{VGA12} & & 
 \end{tabular}
\end{table}
The values are in excellent agreement with the theoretical predictions in both cases. Our measurement of the lifetime of the HFS state in \HBi\ agrees within the uncertainties with the previous measurement by Winter~et~al.~\cite{Win98,Win99} while excluding the older value of Klaft~et~al.~\cite{Kla94}.
\section{Extraction of the $g$-factors}
From the measured transition energies~\cite{Ull17} and lifetimes of the HFS states one can extract the $g$-factors of the bound $1s$ electron for H-like bismuth and the $g$-factor of the bound $2s$ electron for Li-like bismuth.\\
The transition probability between the ground state hyperfine states of \HBi, including first order QED and nuclear corrections, can be written as \cite{Sha98,Bei00}:
\begin{equation}
  \frac{1}{\tau} = \frac{\alpha}{3\hbar} \frac{\Delta E_{\rm HFS}^3}{(m_e c^2)^2} \frac{I}{2I+1} 
                   \left[ g_e - g_I \frac{m_e}{m_p} \right]^2 \; ,
\label{eq:ge}
\end{equation}
where $\Delta E_{\rm HFS}$ is the experimentally determined transition energy of the HFS state, $I$ is the nuclear spin ($I=\frac{9}{2}$ for $^{209}$Bi), $g_e$ is the bound-electron $g$-factor and $g_I$ is the nuclear $g$-factor. The last term can be written as 
\begin{equation}
  g_I\cdot\frac{m_e}{m_p} = \frac{m_e}{m_p\cdot I}\cdot\frac{\mu_{\rm Bi}}{\mu_N}
\end{equation}
with $\mu_{\rm Bi}$ being the magnetic moment of $^{209}$Bi and $\mu_N$ the nuclear magneton. This magnetic moment has recently been re-measured~\cite{Skr18} yielding a value of
\begin{equation}
 \mu_{\rm Bi}=(4.092 \pm 0.002)\cdot~\mu_N \; .
\end{equation}
Together with the experimental HFS splitting of $\Delta E_{\rm HFS}($\HBi$) = 5.08503(2)(9)$\,eV~\cite{Ull17} this leads to a $g_e$-factor for the 1s electron in H-like bismuth of
\begin{equation}
  g_e^{\rm exp}(^{209}{\rm Bi}^{82+}) = 1.7294 \pm 0.0035_{\rm stat.} \pm 0.0019_{\rm sys.}
\end{equation}
in good agreement to the theoretical calculation performed in \cite{MOS04} which yields
\begin{equation}
 g_e^{\rm theo}(^{209}{\rm Bi}^{82+}) = 1.731014 \pm 0.000001 \; .
\end{equation}
Equation~\ref{eq:ge} can also be applied for the calculation of the bound electron $g$-factor in Li-like ions from the experimentally determined lifetime and energy splitting.
With the experimentally determined transition energy of $\Delta E_{\rm HFS}($\LiBi$) = 0.797645(4)(14)$\,eV~\cite{Ull17} and the lifetime presented in this work, the experimentally determined $g_e$-factor for the 2s electron in lithium-like bismuth is found to be:
\begin{equation}
  g_e^{\rm exp}(^{209}{\rm Bi}^{80+}) = 1.928 \pm 0.012_{\rm stat.} \pm 0.010_{\rm sys.} \; ,
\end{equation}
in good agreement to the theoretical value determined from \cite{Mos08}
\begin{equation}
 g_e^{\rm theo}(^{209}{\rm Bi}^{80+}) = 1.934739\pm0.000003 \; .
\end{equation}
\section{Summary}
The lifetimes of the upper HFS states in \HBi and, for the first time, in \LiBi have been measured by the LIBELLE experiment and are compared to previous experimental and/or theoretical results. 
Except for an older measurement of the lifetime in H-like bismuth by Klaft et al.~\cite{Kla94} all results agree within their respective uncertainties.
Using the lifetimes extracted in this work and the energy splittings derived from the same experiment by Ullmann et al.~\cite{Ull17} it is possible to calculate $g$-factors of the bound $1s$ electron for H-like bismuth and of the bound $2s$ electron for Li-like bismuth. Also here the results obtained are in good agreement with theoretical predictions. \\
In the near future, it is forseen to extend the measurements of the specific difference of the HFS transition energies as well as the lifetimes and $g$-factors to the isotopic chain in bismuth. A first candidate for these investigations will be $^{208}$Bi, whose nuclear magnetic moment has been recently measured~\cite{Sch18}.
\section*{Acknowledgments}
This work was supported by the German Federal Ministry of Education and Research (BMBF) under grant numbers 05P12PMFAE, 05P15PMFAA, 05P12RDFA4, 05P12R6FAN and 05P15RGFAA. 
The work was further supported by the Helmholtz International Centre for FAIR (HIC for FAIR) within the LOEWE program by the federal state Hessen. M.L., C.T. and J.U. acknowledge support from HGS-HiRe.\\
We appreciate technical support during the beam time by F.~Nolden, C.~Dimopoulou, J.~Rossbach, and D.~Winters and the GSI accelerator department. We also thank S.~Minami and N.~Kurz for their support in the development of the data acquisition system and R.~J\"ohren for his earlier contributions to the data acquisition and the mirror system. 
\section*{References}
{}
\end{document}